\documentclass{he_symp}
\usepackage{psfig}

\def\cha{{\sl Chandra}}
\begin{document}
\authorrunning{Weisskopf et~al.}
\titlerunning{The prospects for X-ray polarimetry and its potential use for understanding neutron stars}
\title{The prospects for X-ray polarimetry and its potential use for understanding neutron stars}
\author{M.~C.~Weisskopf\inst{1}, R.~F.~Elsner\inst{2}, D.~Hanna\inst{3}, V.~M.~Kaspi\inst{3}, S.~L.~O'Dell\inst{2}, G.~G.~Pavlov\inst{4}, \and B.~D.~Ramsey\inst{2}}
\institute{NASA Marshall Space Flight Center, VP60, Huntsville, AL 35812, USA
\and NASA Marshall Space Flight Center, VP62, Huntsville, AL 35812, USA
\and McGill University Physics Dept., Rutherford Physics Building, 3600 University St., Montreal, QC  H3A 2T8, Canada
\and  Penn State University, Dept. of Astronomy and Astrophysics, 525 Davey Laboratory, University Park, PA 16802, USA
 }
\maketitle

\begin{abstract}
We review the state of the art for measuring the X-ray polarization of neutron stars.
We discuss how valuable precision measurements of the degree and position angle of polarization as a function of energy and, where relevant, of pulse phase, would 
provide deeper insight into the details of the emission mechanisms. 
We then review the current state of instrumentation and its potential for obtaining relevant data. 
Finally, we conclude our discussion with some opinions as to future directions. 
\end{abstract}

\section{Introduction}

Here we discuss the history and the potential scientific impact of X-ray polarimetry for the study of neutron stars.
Despite major progress in X-ray imaging, spectroscopy, and timing, there have been only modest attempts at X-ray polarimetry. 
Indeed, the last such dedicated experiment, conducted by one of us over three decades ago, had such limited observing time and sensitivity that even $\sim 10\%$ degree of polarization would not have been
detected from some of the brightest X-ray sources in the sky, and statistically-significant X-ray polarization was detected in only one of the brightest celestial X-ray sources, the Crab Nebula.
Radio and optical astronomers use polarimetry extensively to probe the radiation physics and the geometry of sources.
Sensitive X-ray polarimetry promises to reveal unique and crucial information about physical processes and structure of neutron stars (and indeed all classes of X-ray sources). 
X-ray polarimetry remains the last undeveloped tool for the X-ray study of astronomical objects and needs to be properly exploited.

\section{Background\label{s:history}}
\begin{figure}
\centerline{\psfig{file=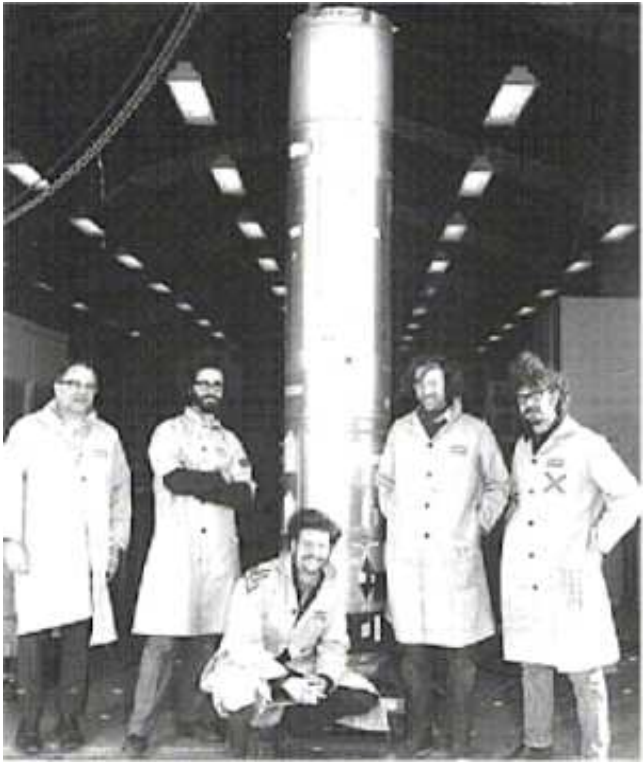,width=8.5cm,clip=} }
\caption{1971 Photograph of the NASA Aerobee-350 sounding rocket \#1709 that first detected polarization from the Crab Nebula. Left to right are R. Novick, 
G. Epstein, M.~C.\ Weisskopf, R. Wolff, \& R. Linke.
\label{f:1709}}
\end{figure}

Only a few experiments have conducted successful X-ray polarimetric observations of cosmic sources.
In rocket observations (Fig.~\ref{f:1709}), the X-ray polarization from the Crab Nebula was measured (Novick et al.\ 1972). 
Using the X-ray polarimeter on the Orbiting Solar Observatory (OSO)-8, Weisskopf et al.\ (1976) confirmed this result with a 19-$\sigma$ detection 
($P =19.2\% \pm 1.0$\%), thus conclusively proving the synchrotron origin of the X-ray emission from this plerionic supernova remnant. 
Unfortunately, because of low sensitivity of those experiments, only upper limits 
were found for polarization from other X-ray sources (e.g., 13.5\% and 60\% for accreting X-ray pulsars Cen X-3 and Her X-1,
respectively; Silver et al.\ 1979).
Since that time, although there have been several missions that have included X-ray polarimeters such as the original {\em Einstein} Observatory, and Spectrum-X, no X-ray polarimeter has actually managed to be launched. 
We discuss this point in more detail in \S \ref{s:conclusion}.

\section{Scientific basis for neutron-star X-ray polarimetry\label{s:science}}

\subsection{Radio pulsars\label{s:rp}}

Radio pulsars are isolated, rotation-powered, neutron stars converting rotational energy to the energy of ultra-relativistic particles and radiation through electromagnetic coupling. 
Strong electric fields and pair production in the very strong (up to a few $\times 10^{13}$ G) magnetic field result in beamed outflow of relativistic particles and radiation and consequent ``search-light'' pulses. 
Theoretical models predict strong linear polarization varying with pulse phase due to the rotation of the neutron star. 
However, details of the emission, as discussed, e.g., in numerous papers presented in this Seminar, and even its location (``polar cap'' versus ``outer gap'') — remain unclear.
X-ray polarimetry could provide decisive information to test detailed models, to determine the emission site, and quite possibly to verify, observationally, the phenomenon of vacuum birefringence as predicted by quantum electrodynamics (QED).

The origin of the high-energy non-thermal pulsar radiation is still a matter of debate.
Controversy remains over the site of this emission: directly above the polar cap, where the coherent radio pulses originate (e.g., Daugherty \& Harding 1992; Harding \& Muslimov 1998), or in the outer magnetosphere (e.g., Cheng, Ho, \& Ruderman 1986a,b; Romani 1996). 
Polarization measurements would discriminate among beaming geometries (e.g., ``polar-cap'' versus ``outer-gap'' models). 

\begin{figure}
\centerline{\psfig{file=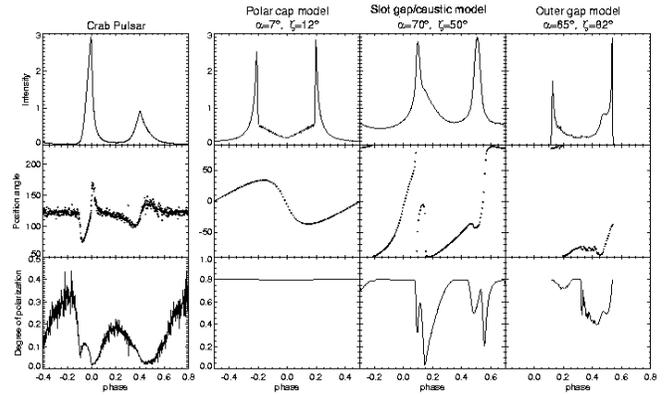,width=9.5cm,clip=} }
\caption{Crab pulsar optical light curve, position angle, and degree of polarization as a function of pulse phase from Kanbach et al.\ 2005. Data are compared to three different predictions of theoretical models. Courtesy A. Harding.
\label{f:optical}}
\end{figure}

The requirements on X-ray polarimetry may be estimated by examining the optical polarimetry of the Crab pulsar (e.g., Smith et al.\ 1988; Romani et al.\ 2001),
which shows (Fig.~\ref{f:optical}) high linear polarization, varying rapidly through each pulse component. 
Because the field line projection determines the polarization position angle, we expect a close, but not necessarily identical, correspondence between the optical and X-ray sweep of the position angle. 
Previous polarimetry of the Crab, limited to a single energy (2.6 keV) could place only upper limits of 20\% to 30\% on the pulsar's polarization in wide phase bins (Silver et al.\ 1978). 
What is needed are much more sensitive measurements capable of providing, at a minimum, data over a large number of pulse phase bins that are small enough to resolve different features of the pulse profile.

The pulsar's X-ray emission is almost certainly synchrotron radiation. 
If, however, as has been proposed (e.g., Sturrock, Petrosian, \& Turk 1975), the optical emission were curvature radiation, the X-ray polarization would be orthogonal to the optical polarization (Fig.~\ref{f:optical}). 
If, instead, the remarkably flat optical spectrum were 
a low-pitch-angle extension of the X-ray synchrotron population (Crusius-Waetzel, Kunzl, \& Lesch 2001), the larger X-ray pitch angle would smooth the position-angle sweep (the variation of the position-angle as a function of pulse phase).
Thus, the X-ray-polarization profile (polarization amplitude and position angle as a function of pulse phase) would be a sensitive probe of the magnetospheric particle distribution over pitch-angles.

Polarimetry also offers an interesting opportunity to observe an exotic QED effect --- vacuum birefringence --- induced by a strong magnetic field. 
Predicted nearly 70 years ago (Euler \& Kockel 1935, Heisenberg \& Euler 1936, Weisskopf 1936), the effect is yet to be verified observationally. 

The effect follows from the result (e.g., Tsai \& Erber 1975) that the indices of refraction for radiation polarized parallel ($n_{\|}$) and perpendicular ($n_{\perp}$) to the plane formed by the direction of propagation and the magnetic field are different and depend on the field strength:

%
%
\begin{eqnarray}
n_{\|} & \approx & 1 + \frac{\alpha}{4\pi} \sin^{2}\theta \left[\frac{14}{45}\left(\frac{B}{B_{\rm cr}}\right)^{2}-\frac{13}{315}\left(\frac {B}{B_{\rm cr}}\right)^{4}\right]\\
n_{\perp} & \approx & 1 + \frac{\alpha}{4\pi} \sin^{2}\theta \left[\frac{8}{45}\left(\frac{B}{B_{\rm cr}}\right)^{2}-\frac{379}{5040}\left(\frac{B}{B_{\rm cr}}\right)^{4}\right]
\end{eqnarray}
for photon energies below the one-photon pair production threshold
and magnetic fields much lower than $B_{\rm cr}=m^2c^3/e\hbar \approx 4.4 \times 10^{13}$\ G (here  $\alpha$ is the fine structure constant and $\theta$ is the angle between the direction of propagation of the photons and the magnetic field).
Thus, for a field of $3 \times 10^{12}$ G, we have $n_{\|}-n_{\perp}\sim 4 \times 10^{-7}$ for propagation transverse to the field lines. 
At 1 keV, the path length for one wave retardation is only a few mm. 

Pavlov, M\'{e}sz\'{a}ros, and co-workers investigated the influence of vacuum birefringence on X radiation from neutron stars (see Pavlov \& Gnedin 1984 and M\'{e}sz\'{a}ros 1992 for reviews). 
To accurately locate the X-ray-emitting site and infer its properties, 
vacuum birefringence effects on radiation propagating in a nonuniform magnetic field must be taken into account.
For instance, if the emission site is near the neutron-star surface (as in polar-cap models), the vacuum birefringence leads to an energy dependence of the polarization direction at a particular rotation phase (Heyl \& Shaviv 2000).
This results in a $\sim10^\circ$ phase shift between the optical and X-ray polarization swings, with the X-ray sweep leading. 


The measurement of such a phase shift would not only locate the emission site, but it would also represent a direct observational manifestation of vacuum birefringence.

\subsection{Magnetars}

Soft Gamma-ray Repeaters (SGRs) and Anomalous X-ray Pulsars (AXPs) are presumably isolated, magnetic-powered neutron stars, converting magnetic energy ultimately into high-energy radiation. 
SGRs and AXPs are likely to be magnetars, i.e. neutron stars with extremely strong ($10^{14-15}$ G) magnetic fields. 
Magnetically coupled seismic activity possibly results in high-energy radiation and plasma outflows, occasionally in extremely luminous (up to $10^{47}$ erg s$^{-1}$) giant flares of SGRs. 
Radiation emitted in such superstrong magnetic fields is inevitably highly polarized (e.g., Niemiec \& Bulik 2006, and references therein).
X-ray polarimetry can provide important data for understanding the nature of magnetars and for studying physical processes in extreme magnetic fields.

In the widely accepted magnetar model (Duncan \& Thompson 1992), the neutron 
star's strong magnetic field powers persistent emission through low-level seismic activity and heating of the stellar interior; it powers the burst emission through large-scale crust fracture (Thompson \& Duncan 1995, 1996). 
However, there is no generally accepted detailed model for the SGR emission, particularly in the active (burst) phase (Lenters et al.\ 2003, and references therein), with peak soft-X-ray luminosities between $10^{38}$ and $10^{44}$ erg s$^{-1}$ (Hurley 2000). 
Sources such as 1806--20 may have even brighter soft components during giant flares.
The persistent radiation of magnetars is relatively faint in soft X-rays ($L_x \sim 10^{34-35}$ erg s$^{-1}$ in the {\sl Chandra}-{\sl XMM} band).
However, recent observations with {\sl INTEGRAL} and {\sl RXTE} have revealed
hard-X-ray tails in the magnetar spectra (e.g., Kuiper, Hermsen, \& 
Mendez 2004; G\"{o}tz et al.\ 2006), with 20--100 keV luminosities up to $\sim 10^{36}$ erg s$^{-1}$, which makes this range promising for polarization observations.
If an SGR becomes active, the polarization will be amenable to measurement. 
Giant flares are too rare and brief to easily observe and might saturate many instruments.
Still one can expect to observe an SGR during an active period when it produces numerous short (1-sec) bursts with a flux-dependent event frequency --- $dN/dS \propto S^{-5/3}$ (G\"{o}\~{g}\"{u}\c{s} et al.\ 2001). 
For activity such as SGR 1900+14 exhibited in 1998 August or in 2001 April, there would be about 30 short bursts, with burst fluence $10^{-7}$--$10^{-5}$ erg cm$^{-2}$ (25--100 keV band) in a time interval of 100 ks.
It is not inconceivable to be able to detect polarization from the total fluence under these conditions. 

\subsection{DINSs and CCOs}

``Dim'' isolated neutron stars (DINSs) are radio-quiet and non-accreting, exhibiting predominately thermal emission ($kT\approx 50$--100 eV) from the neutron-star surface. 
Periods in the range of 3--12 s have been measured for five of of the seven
currently known DINSs (Haberl 2006 and references therein), and for two of them period derivatives have been also measured, which allows one to estimate the dipole components of magnetic field, $B=2.4$ and $3.4\times 10^{13}$ G (Kaplan \& van Kerkwijk 2005a,b), approaching the superstrong magnetic fields of magnetars.
Although the spectrum of the brightest DINS, RX J1856.4$-$3754 (for which no pulsations have been found even in very long exposures) is close
to a perfect blackbody, most of these objects show puzzling absorption lines in their spectra, whose origin has not been understood yet (van Kerkwijk \& Kaplan 2006; Haberl 2006). 
Most likely, these lines are formed in Hydrogen or Helium atmospheres of the neutron stars, but the actual atomic transitions involved, and even the chemical composition of the atmospheres, remain to be understood. 
Since transitions between different types of atomic states (so-called tightly-bound and hydrogen-like states) are sensitive to different (mutually orthogonal) polarizations, polarization measurements would be very helpful in understanding the type of the transitions involved, which, in turn, would establish the chemical composition and the strength and geometry of the magnetic field (Pavlov \& Bezchastnov 2005).
Moreover, even the continuum spectrum of neutron stars should be strongly polarized (typically, a few $\times 10\%$) because the atmospheric opacities are
very different in different polarizations (e.g., Pavlov et al.\ 1995), and polarization degree and position angle show strong variations with pulsar rotation phase (Pavlov \& Zavlin 2000).
Therefore, using a soft-X-ray polarimeter, one could resolve the puzzle of RX J1856.4$-$3754 --- its period could easier be found in polarized light, and the polarization variations could provide useful information about its (currently unknown) magnetic field and explain the lack of spectral features. 
Finally, since the spectra and light curves of polarization of thermal radiation of neutron stars bear unique signatures of the vacuum polarization in a strong magnetic field (van Adelsberg \& Lai 2006 and references therein), polarization
observations of DINSs could not only detect this effect but also use it for investigations of surface layers of neutron stars.

We also note that the same arguments are applicable to another class of radio-quiet neutron stars --- the central compact objects (CCOs) in supernova remnants (see Pavlov et al.\ 2002, 2004 for a review). 
These objects also show thermal spectra, with temperatures in the range of 100--500 eV (hotter than DINSs but somewhat colder than magnetars), and they show neither pulsar activity  (e.g., pulsar-wind nebulae or $\gamma$-ray emission) nor magnetar behavior (e.g., bursts). 
Their nature is even less understood than that of magnetars and DINSs.
For instance, the CCO in the Cas A SNR, discovered in the first-light {\sl Chandra} observations (Tananbaum 1999), shows a thermal-like spectrum emitted from a small fraction of the neutron star surface, similar to magnetars (Pavlov et al.\ 2000), but no pulsations (Chakrabarty et al.\ 2001). 
A particularly interesting member of this class is 1E 1207.4$-$5209 in the PKS 1209$-$51/52 SNR, the only confirmed pulsator among the CCOs (Zavlin et al.\ 2000), and the only CCO whose spectrum shows at least two absorption lines, at 0.7 and 1.4 keV (Sanwal et al.\ 2002).
The origin of the lines remains unknown.
Sanwal et al.\ (2002) have concluded that these lines cannot be associated with transitions in Hydrogen atoms and argued that neither electron nor proton cyclotron resonance could cause these features. 
These authors suggest that the lines could due to absorption by once-ionized Helium in a magnetic field $B\sim 2\times 10^{14}$ G (see also Pavlov \& Bezchastnov 2005), while Mori \& Hailey (2006) argue that the lines could be formed in an Oxygen atmosphere with $B\sim 10^{11-12}$ G. 
Whatever is the origin of the lines and the small, hot emitting areas in CCOs, 
only high magnetic fields, possibly with strong multipole components, could explain their properties. 
This means that the CCO's radiation is inevitably strongly polarized, and, similar to DINSs, polarization observations would be extremely useful for solving the puzzles of these unusual neutron stars.

\subsection{Pulsating X-ray binaries}

Pulsating X-ray binaries are accretion-fed neutron stars, converting kinetic energy into X-ray emission at the stellar surface. 
Rotation and accretion-flow anisotropy, induced by very strong magnetic fields ($10^{12}$ to $10^{13}$ G), modulate the X rays. 
Most theoretical models predict that the linear polarization of this X radiation is high and varies with pulse phase (due to rotation of the star) and also varies with energy (due to energy-dependent opacity, cyclotron resonance, and vacuum birefringence). 
X-ray polarimetry would provide crucial information to test detailed models, to infer parameters and geometries, and to verify vacuum birefringence observationally.

More than 50 binary X-ray sources in our Galaxy and the Magellanic Clouds exhibit pulses with periods from 69 ms to 23 min (e.g., Nagase 1989; Bildsten et al.\ 1997).
Interpreting absorption features between 10 and 100 keV (Coburn et al.\ 2002, and references therein), observed in about a dozen pulsating X-ray binaries, as cyclotron absorption lines (Gnedin \& Sunyaev 1974) implies very strong magnetic fields, $\sim 10^{12-13}$ G, strengths that we commonly associate with these objects.
Under such conditions, X-ray emission, absorption, and scattering depend strongly on energy, direction, and polarization.

Detailed theoretical studies (M\.{e}sz\'{a}ros et al.\ 1988; M\.{e}sz\'{a}ros 1992, and references therein) show that the linear polarization depends strongly upon the geometry of the emission region (accretion column), varies with energy and pulse phase, and reaches values as high as 60\% to 70\% for favorable orientations.  
Calculating the X-ray spectrum, pulse profile, and polarization from a high-temperature, strongly magnetized, rotating neutron star is complex. 
Further, the results depend strongly upon the assumed distribution of magnetic field, temperature, and density in the accretion column (Lamb 1977; M\.{e}sz\'{a}ros 1982; Kaminker, Pavlov, \& Shibanov 1982, 1983; Arons, Klein, \& Lea 1987; Brainerd \& M\.{e}sz\'{a}ros 1991; Isenberg, Lamb, \& Wang 1998; Araya-G\.{o}ches \& Harding 2000).
Nevertheless, theoretical modeling has now progressed to the point that X-ray polarization measurements can test models and infer parameters of the accreting matter and of the neutron star.

For example, phase-resolved polarimetry can distinguish between “pencil” and “fan” radiation patterns, corresponding to different emission-region geometries. 
Because the degree of linear polarization is maximum for emission perpendicular to the magnetic field, the flux and degree of polarization are in-phase for fan beams, but out-of-phase for pencil beams. 
Particularly interesting are those cases (e.g., Her X-1, GX1+4, and 4U1626$-$67) when pulse profiles change dramatically with energy, including pulse-maxima reversals between 1 and 20 keV (White, Swank, \& Holt 1983). 
Several authors (Nagel 1981a,b; White, Swank, \& Holt 1983) believe that such behavior requires both fan and pencil beam components, with each component dominating at different energies. 
Hence, polarimetry can differentiate among the semi-empirical models (e.g., M\.{e}sz\'{a}ros \& Nagel 1985; Dermer \& Sturner 1991; Bulik et al.\ 1992; Isenberg  Lamb, \& Wang 1998) that qualitatively reproduce the pulse profiles but predict quite different phase dependences for the linear polarization.

Because the linear-polarization direction lies either parallel or perpendicular to the magnetic field (depending upon photon energy and absorption depth), the sweep of the polarization position angle with pulse phase specifies the magnetic-field geometry.
For instance, abrupt position-angle changes would indicate a non-dipolar field (Elsner \& Lamb 1976).
If observed, these position-angle changes would support other evidence for such fields in some accreting pulsars (Bulik et al.\ 1992), due perhaps to thermo-magnetic effects (Blandford, Applegate, \& Hernquist 1983) or crustal breaking and migration of field-carrying platelets (Ruderman 1991). 
Such measurements require a polarimeter sensitive in the energy bands near the electron-cyclotron energy $E_{ec}= (11.6\, {\rm keV}) (B/10^{12}$G). 
Because the polarization dependence on energy is strongest near $E_{ec}$, one could establish which model is most reliable and obtain magnetic-field measurements for sources in which the cyclotron line is yet undetected.

As with radio pulsars (\S \ref{s:rp}), X-ray polarimetry of pulsating X-ray binaries may detect effects of vacuum birefringence. 
Recent studies of neutron-star atmospheres (e.g., Lai \& Ho 2003) and magnetospheres (e.g., Heyl \& Shaviv 2000) treat this phenomenon.
The most vivid polarization signature is a $90^\circ$ position-angle jump at an energy-dependent phase, occurring where normal-mode propagation through the so-called ``vacuum resonance'' (Pavlov \& Shibanov 1979) changes from adiabatic to nonadiabatic (Pavlov \& Gnedin 1984; Lai \& Ho 2003).
Detection of such a jump would provide a direct observation of this QED effect. 
Moreover, the jump's phase at a given energy depends on accretion-column inclination and density scale length in the radiating region, affording estimates of these quantities.

In the only X-ray polarimetry on pulsating X-ray binaries to date, Silver et al.\ (1979) found 99\%-confidence upper limits of 13.5\% polarization for Cen X-3 and 60\% polarization for Her X-1, at 2.6 keV. 
In order to make significant progress one needs dramatic improvements in sensitivity whereby the polarization may be studied both as a function of energy and as a function of pulse phase. 

\subsection{Other applications}

We have concentrated on the role that X-ray polarization measurements can play in understanding the X-ray emission from neutron stars. 
It is worth emphasizing that X-ray polarimetry has far broader applications and would allow one to explore such systems as Galactic accretion-disks, Galactic superluminal sources, active galactic nuclei, etc.

Galactic accretion-disk systems involve accretion-powered neutron stars or black holes, converting kinetic energy into X-ray emission in the hot inner regions of the disk. 
While the X-ray polarization of radio pulsars, magnetars, and pulsating X-ray binaries is due to strong neutron-star magnetic fields, the polarization of accreting binaries with a low-field-neutron-star or black-hole primary will likely be dominated by scattering.
Due to their complexity, accretion-disk systems as a group exhibit rich diversity: magnetodisks, coronae, winds, quasi-periodic oscillations, 
millisecond pulsations in spun-up pulsars, bursting, etc. 
X-ray polarimetry can probe the properties of the complex structure of accretion-disk systems, and explore the space-time structure close to a black hole. 
This latter is an especially interesting application of X-ray polarimetry.

Galactic superluminal sources (microquasars) and extragalactic sources such as 
AGNs (quasars, blazars, Seyfert galaxies, etc) are all disk-jet sources, converting kinetic energy of accreted material into X radiation and directed beams of relativistic plasma. 
Such sources are comprised of an interacting binary containing a black hole, stellar-mass size in the case of microquasars and supermassive for the others. X-ray polarimetry can provide important information on the X-ray emission mechanism and the site  (disk, corona, or jet) of its origin.

\section{Instrumental approaches\label{s:instrument}}

There are a limited number of ways to measure linear polarization in the range 0.1--50~keV, sufficiently sensitive for astronomical sources.
Before reviewing some of these, we emphasize that {\em meaningful X-ray polarimetry of such sources is difficult}: 
\begin {enumerate}
\item In general, we do {\em not} expect sources to be strongly ($\gg$10\%) polarized.
For example, the maximum polarization from scattering in an optically think, geometrically thin, accretion disc is only about  10\% at the most favorable (edge-on) viewing angle.
Hence, most of the X rays from such a source carry no polarization information and thus merely increase the background (noise) in the polarization measurement.
\item With one notable exception~--- namly, the Bragg-crystal polarimeter (\S \ref{s:bragg})~--- the modulation of the polarization signal, which is the signature of polarization in the detector, is much less than 100\% (typically, 20\%--40\%) even for completely polarized source.
Unfortunately, a Bragg-crystal polarimeter has but a narrow spectral response, thus limiting the number of photons detected and providing little information on the spectral dependence of the polarization.
\item The degree of linear polarization is positive definite, so that any polarimeter will always measure (not necessarily statistically significantly) a polarization signal, even from an unpolarized source.
Consequently, the statistical analysis (\S \ref{s:statistics}) becomes somewhat complicated.
\end {enumerate}

It is partly for these reasons that X-ray polarimetry has not progressed as rapidly as X-ray imaging, timing, and spectroscopy, since the pioneering experiments performed in the early 1970's.
There are also sociological and psychological reasons, especially those involving the competition for observing time and the projected rate of return for instruments at the focus of telescope facilities (see also \S \ref{s:conclusion}) which have played a role in stifling the development of X-ray polarimetry.

Two different types of X-ray polarimeters have flown to date~---
Bragg-crystal polarimeters (\S \ref{s:bragg}) and scattering polarimeters (\S \ref{s:scattering}).
Note that we here differentiate between instruments that have been expressly designed and constructed to measure polarization and those that possess a degree of sensitivity to polarization, but were not designed for this purpose.
We shall comment on these latter in \S \ref{s:conclusion}.
In this paper, we also discuss (\S \ref{s:photoelectron}) the advantages and disadvantages of a more ``modern'' approach to studying X-ray polarization, which uses the polarization dependence of K-shell photoelectron emission.

We emphasize the importance of the comparison we make here, as there appears to be some confusion concerning the relative merits of the different approaches.
The recent literature has asserted such statements as ``conventional polarimeters based on Bragg diffraction or Thompson scattering methods are characterized by a poor sensitivity...'' (Bellazzini et al.\ 2006).
Such broad statements are misleading, if not incorrect, in that they ignore the various contexts in which an X-ray polarimeter might fly, as well as issues of proven performance, cost, and simplicity.

\subsection{Polarimeter basics}

All the polarimeters we discuss here have the following characteristic in common.
The detected polarization signal behaves as
\begin{equation}
S = \bar{S}[1 + a_{0}\cos(2\psi+\phi_0)], \label{e:1}
\end{equation}
where $\psi$ is an angle with respect to the instrument's axis, in the plane transverse to the incident photon's direction.
Here $a_0$ and $\phi_0$ are related to the degree of linear polarization and its position angle, respectively.

\subsection{Statistics\label{s:statistics}}

We assume that the detected signal is drawn from a broad-band noise source characterized by a mean $\bar{S}$ and variance $\sigma^2$.
Then the probability of measuring a particular amplitude of modulation $a$ and phase $\phi$ is given by 
\begin{equation}
P(a,\phi) = \frac{N \bar{S}^{2} a}{4\pi\sigma^{2}}\exp\left[ - \frac{N \bar{S}^{2}}{4\sigma^{2}}(a^2 + a_{0}^{2} - 2aa_{0}\cos(\Delta\phi)\right], \label{e:2}
\end{equation} 
where $\Delta\phi \equiv \phi-\phi_0$ and $N$ is the number of different values of $\psi$ for which measurements were made~---i.e. the number of data points.

It follows that the probability of measuring a particular amplitude $a$ independent of $\phi$ is 
\begin{equation}
P(a) = \frac{N\bar{S}^{2}a}{2\sigma^2} \exp \left[ -\frac{N\bar{S}^2}{4\sigma^2}(a^2+a_{0}^{2}) \right] I_0(\frac{N\bar{S}^{2}a a_0}{2\sigma^2}), \label{e:3}
\end{equation} 
where $I_0$ is the modified Bessel function of order zero.

The probability of measuring a particular angle $\phi$ independent of the amplitude $a$ is:
\vspace{0.1in}

\noindent $P(\phi) = \frac{1}{2\pi}\exp(-\frac{N\bar{S}^{2}a_0^2}{4\sigma^2}) + (\frac{N}{2})^{1/2} \frac{a_0\bar{S} \cos(\Delta\phi)}{2\pi\sigma}$ 

\begin{equation}
~\times \exp~[-\frac{N^2\bar{S}^2 \sin^2 \Delta\phi}{4\sigma^2}] \int_{-\infty}^{(N/2)^{1/2}\frac{a_0\bar{S}\cos\Delta\phi}{\sigma}} \exp(-\frac{u^2}{2})du. \label{e:4}
\end{equation} 

In the following we assume Poisson distributed data and set $\sigma^2=\bar{S}$.
There are two interesting limiting cases which can be calculated analytically. In the first we consider large arguments of the Bessel function in Eq.~\ref{e:3} and for $a$ close to $a_0$; P(a) then becomes a normal distribution with $\sigma_a=(2/N)^{1/2}$. 
Similarly, when the upper limit of the integral in Eq.~\ref{e:4} gets very large compared to 1 and for $\phi$ close to $\phi_0$, P($\phi$) becomes a normal distribution with $\sigma_\phi=\sigma_a/a_0$.

To establish an instrument's sensitivity to polarized flux, the most relevant statistical question is, if the data are unmodulated (no real measure of polarization: $a_0=0$), what is the probability of measuring, by chance, an amplitude of modulation that is greater than or equal to the measured value?
The amplitude of modulation is, after all, positive definite and a value will be measured.
In this case, Eq.~\ref{e:2} may be integrated analytically and, if the data are Poisson distributed, one finds

\begin{equation}
P(a' \geq a) =  \int_{a}^{\inf} P(a')~da' = \exp~(-\frac{N\bar{S}a^2}{4}). \label{e:5}
\end{equation} 

\noindent Note that $N\bar{S}$ is simply the total number of counts.
It has become customary to single out the amplitude that has only a $1\%$ probability of chance occurence.
Solving Eq~\ref{e:5}, this amplitude ($a_{1\%}$) is
\begin{equation}
a_{1\%}=\frac{4.29} {(N\bar{S})^{1/2}}. \label{e:6}
\end{equation} 

The total number of counts, $\bar{S}$, is simply related to the source ($R_S$) and background ($R_B$) counting rates and the total observing time (T) through $N\bar{S} = (R_S + R_B)T$.
Furthermore, we are interested in the modulation expressed as a fraction of the mean {\em source} counts, not the mean total counts, i.e. $a_S = a_{1\%}/\bar{S}$ so that

\begin{equation}
a_{S}=\frac{4.29}{R_S}\left[ \frac{R_S+R_B}{T}\right] ^{1/2}. \label{e:7}
\end{equation} 

Finally, one needs to account for the possibility that the polarimeter does not fully respond to $100$\%-polarized radiation. 
It is convenient to introduce the ``modulation factor'', $M$, which is the degree of modulation expected in the absence of background and for a $100$\%-polarized beam. 
Thus, independent of the position angle, the minimal detectable polarization at the $99$\% confidence level, ${\rm MDP}_{99}$, is

\begin{equation}
{\rm MDP}_{99}=\frac{a_S}{M}=\frac{4.29}{MR_S} [\frac{R_S+R_B}{T}]^{1/2}. \label{e:8}
\end{equation} 

It is sometimes mistakenly assumed that Eq.~\ref{e:8} for the minimal detectable polarization describes the uncertainty of a measurement of the polarization: {\em That is not the case.}
Eq.~\ref{e:8} indicates when one may be confident that the signature of polarization has been detected~--- i.e., that the source is {\em not unpolarized}~--- but not the uncertainty of its value (Eq.~\ref{e:2}).
We emphasize this point because the minimal detectable polarization (MDP) often serves as {\em the} figure of merit for polarimetry.
While it is {\em a} figure of merit that is useful and meaningful, a polarimeter useful for attacking astrophysical problems must have an MDP significantly smaller than the degree of polarization to be measured.

\subsection{Crystal polarimeters}\label{s:bragg}

The first successful X-ray polarimeter for astronomical application utilized the polarization dependence of Bragg reflection.
Weisskopf et al.\ (1972) describe the first sounding-rocket experiment (Fig.~\ref{f:1709}) using crystal polarimeters, which Schnopper \& Kalata (1969) had first suggested for an astronomical application.

To understand the operating principle of such devices, consider a single flat crystal.
The number of reflected X-rays (N) during an observation of length T, given incident radiation with a spectral distribution I(E) (keV/keV/cm$^2$-sec), is 
\begin{equation}
\frac{N}{T} = \int_{0}^{\infty}\frac{I(E')}{E'} R(E',\theta)A(\theta)dE',
\end{equation} 
where $A(\theta)$ is the projected area of the crystal in the direction of the incident flux and $R(E,\theta)$ is the probability that a photon of energy E incident on the crystal at angle $\theta$ will be reflected.
For a continuum it can be shown (see, e.g., Angel \& Weisskopf 1970) that this expression reduces to
\begin{equation}
\frac{N}{T} = I(E) A(\theta_{B})\bigtriangleup\theta(E) cot(\theta_{B}),
\end{equation}
where $E$ is related to $\theta_{B}$ through the Bragg condition:
\begin{equation}
E = \frac{nhc}{2d \sin(\theta_{B})}.
\end{equation} 
Here $d$ is the interplanar spacing of the crystal lattice, $n$ is the order of the reflection, and $\bigtriangleup \theta (E)$ is the integrated reflectivity at incident energy $E$
\begin{equation}
\bigtriangleup\theta (E) = \int R(E,\theta) d\theta.
\end{equation} 

For partially polarized radiation ($P\leq1.0$) 
\vspace{0.1in}

\noindent $\bigtriangleup\theta (E) = \frac{N_{s}^{2}F^{2}r_{0}^{2}} {2\mu(E)} (\frac{hc}{En})^{3}$
\begin{equation}
 ~\times(\frac{1}{\sin2\theta_B}-\frac{\sin2\theta_B}{2}(1+P\cos2\phi)),
\end{equation} 
where $\phi$ is the angle between the electric vector and the plane of reflection, and $N_s$ is the number of scattering cells per unit volume, $F$ is the crystal structure factor, $r_{0}$ is the classical electron radius, and $\mu$ is the absorption coefficient.
The variation of the counting rate as a function of $\phi$ is maximal for $\theta_{B}$ at 45 degrees and the azimuthal variation goes as $\cos2\phi$.

\begin{figure}
\centerline{\psfig{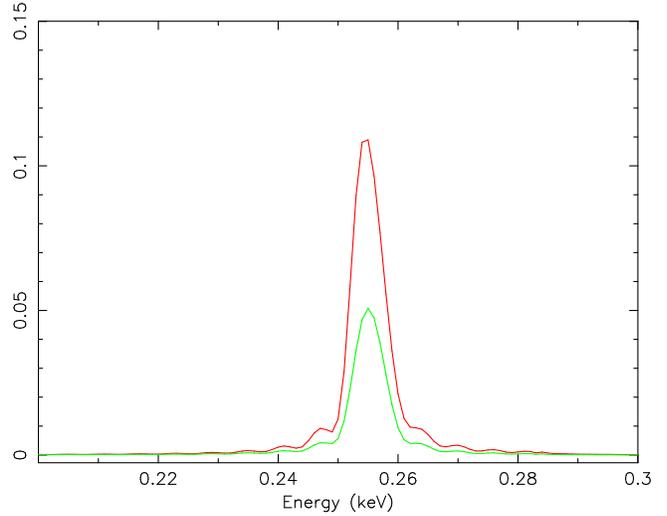} }
\caption{Theoretical reflectivity versus energy for a 40 bilayer, Ni/C synthetic multilayer, each layer being 34 \AA thick. Fractional thickness Ni (0.4) and C (0.6). For 0 \AA (upper) and 5 \AA (lower) interlayer roughness.
\label{f:plexasr}}
\end{figure}

The integrated reflectivity is not the same for all crystals, even of a given material, but depends on the relative orientation of the crystal domains. 
These latter may be viewed as small ``crystalets''.
The integrated reflectivity is highest in the case of the ``ideally imperfect'' or ``mosaic'' crystal where perfect alignment of the the crystal planes is maintained only over microscopic domains in three dimensions.
If these domains are much less than an absorption length in depth along the direction of the incident photon, then an X-ray entering the crystal may encounter many such domains, each at a slightly different Bragg angle, enhancing the probability of a Bragg reflection taking place before the photon might be absorbed.
One can contrast this behavior with that which takes place in a perfect crystal where there is (essentially) only one very large domain with a single orientation; only X-rays with a very narrow bandwidth ($<< 1$ eV) can satisfy the Bragg condition, and all other X-rays are absorbed (or continue to pass through the crystal).
As a consequence, perfect crystals have very low integrated reflectivity, which makes them poor candidates for polarization analyzers of the continuum fluxes prevalent from astrophysical sources.
An ideally imperfect crystal can have an integrated reflectivity 10 to 100 times greater than that of an ideally perfect crystal of the same material.
Angel \& Weisskopf (1970) performed a theoretical study of the integrated reflectivity of a number of naturally occurring crystals and discussed their potential for X-ray astronomy applications.
The highest integrated reflectivity they found was for graphite ($\bigtriangleup\theta (E) = 1.5 \times 10^{-3}$).
Actual realizations using pyrolitic graphite have achieved values closer to $1  \times 10^{-3}$.
Synthetic multilayer crystals, wherein alternating layers of high-Z, low-Z materials (e.g., Ni/C) are constructed, may achieve comparable and even larger integrated reflectivities at low energies. 
The performance of these crystals depends critically on the inter-layer surface roughness which is not easy to control.
Figure ~\ref{f:plexasr} e.g., illustrates such effects.
Multilayer crystals operating at low energies are especially attractive for observing effects from the so-call ``dim'' (they are anything but dim) isolated neutron stars.

Only three crystal polarimeters have ever been constructed for extra-solar X-ray applications and only two --- both using graphite crystals without X-ray telescopes --- were ever flown (sounding rocket, Weisskopf et al.\ 1972; OSO-8 satellite, Weisskopf et al.\ 1976; Spectrum-X (not flown), Kaaret et al.\ 1994 and numerous references therein.)

One of the strongest virtues of the crystal polarimeter is, for Bragg angles near 45 degrees, that the modulation of the reflected flux approaches 100\%. 
One can see from Eq~\ref{e:8} that this is very powerful {\em all other things being equal}.
Thus a factor of two increase in the modulation factor improves the minimum detectable polarization (MDP) by a factor of 2. 
To achieve the same improvement in sensitivity by other means would require either an increase in effective area or observing time by a factor of 4.

The most severe disadvantage of the crystal polarimeter is the narrow bandwith  of the response --- about 23 eV for graphite with a mosaic spread (rocking curve width) of $0.5\deg$.
The integrated reflectivity from the second order Bragg reflection is smaller than that from the first order and, of course, the typical flux from astronomical sources are usually comparatively weaker so that the overall loss in sensitivity renders the second (and higher) order(s) of marginal utility.
Filling in gaps in energy coverage therefore requires using different crystals, which, in general, implies a very poor ``filling factor''.
Here the filling factor refers to one's ability to make use of the real estate in a satellite payload that lies perpendicular to the incident flux.
Unless stacked (and, because of photoelectric absorption, stacking cannot be extended arbitrarily) two crystal polarimeters, which effectively cover two energies, divide the available area in half, three --- one-third, etc.
This may be contrasted to the scattering and electron tracking polarimeters discussed below which cover a much larger bandwidth with a filling factor of unity, typically, however, at the price of a smaller modulation factor.

\subsection{Scattering polarimeters}\label{s:scattering}

There are two scattering processes from bound electrons that must be considered: coherent and incoherent scattering. 
A comprehensive discussion of both of these processes may be found in many atomic physics textbooks (see, e.g., James 1965).
Coherent scattering dominates at small scattering angles.
In the limit of zero scattering angle, the X-ray behaves as if it were scattered from a charge $Ze$, where $e$ is the charge of an electron. 
Coherent scattering, therefore, leads to an enhancement of forward scattering over pure Thomson scattering from free electrons.
In the non-relativistic limit, the cross-section for coherent scattering for X-rays traveling along the $z$-axis and polarized along the $y$-axis is
\begin{equation} 
\frac{d\sigma_{coh}}{d\omega} = r_0^{2}[\cos^2\theta\cos^2\phi+\sin^2\phi] |F|^2.
\end{equation}
Here $r_0$ is the classical electron radius, $\theta$ is the polar scattering angle, and $\phi$ is the azimuthal angle measured from the x-axis. 
Tables of the form factor F may be found in the literature (e.g., Hansen et al.\ 1964)

Incoherent scattering dominates at larger scattering angles and approaches the Thomson limit at sufficiently large angles.
In the non-relativistic limit, the cross-section for incoherent scattering of X-rays polarized along the x-axis is:
\begin{equation} 
\frac{d\sigma_{incoh}}{d\omega} = r_0^{2}[\cos^2\theta\cos^2\phi+\sin^2\phi] I.
\end{equation}
Tables for the incoherent scattering function, I, are also available in the literature (e.g., by Cromer \& Mann 1967).

Various factors dominate the consideration of the design of a scattering polarimeter.
The most important of these are: (1) to scatter as large a fraction of the incident flux as possible while avoiding multiple scatterings (which clearly blurs the polarization dependence); (2) to achieve as large a modulation factor as possible; (3) to collect as many of the scattered X-rays as possible; and (4) to minimize the detector background.
The scattering competes with photoelectric absorption in the material, both on the way in and, of course, on the way out.
The collection efficiency competes with the desire to minimize the background.
Most practical designs have the detector integrating the two scattering angles over some range which impacts the modulation factor.

Only two polarimeters of this type have ever been constructed for extra-solar X-ray applications and only one - utilizing blocks of lithium with proportional counters covering the 4 sides of the blocks orthogonal to the incident flux - was ever flown (rockets - three times: in 1968, see Angel et al.\ 1969; in 1969 see Wolff et al.\ 1970, and in 1971 see, e.g., Novick et al.\ 1972; satellite - Spectrum-X (never flown) see Kaaret et al.\ 1994 and numerous references therein.)

The virtue of the scattering polarimeter is that it has reasonable relative efficiency over a moderately large energy bandwidth, typically several keV in width.
The bandwidth is large compared to the energy resolution of potential detectors, e.g., proportional counters, CCDs, etc., so that polarization measurements as a function of energy are feasible.
The principal disadvantage is a modulation factor less than unity, since only for scattering into $90\deg$ will the modulation approach unity in the absence of background and for a $100$\%-polarized beam. 
In order to obtain any reasonable efficiency requires integrating over a range of scattering angles and realistic modulation factors are under 50\%, unless the device is placed at the focus of a telescope (the modulation factor for the scattering polarimeter on Spectrum-X reached $\sim$75\%) where it is feasible to make the scattering volume small which then limits the range of possible scattering angles.

The two most popular materials that have been considered for scattering polarimeters are lithium and beryllium.
The lower the Z, the lower the peak response energy, and, for cosmic X-ray sources, the higher the sensitivity.
The K-absorption edges for these two materials are at .0554 keV (Li) and 0.188 keV (Be).
The peak energy response of typical practical designs are typically $\simeq 7$ keV (Li) and $\simeq 15$ keV (Be) but it should be noted that the exact peak energies are somewhat design/detector-dependent.

\subsection{Photoelectron tracking polarimeters} \label{s:photoelectron}

The angular distribution (see, e.g., Heitler 1954) of the K-shell photoelectron emitted as a result of the photoelectric absorption process depends on the polarization of the incident photon. 
In the non-relativistic limit
\begin{equation} 
\frac{d\sigma}{d\Omega} =f(\zeta) r_{0}^{2}Z^5\alpha_0^4(\frac{1}{\beta})^{7/2}4\sqrt{2} \sin^{2}\theta \cos^{2}\phi.
\end{equation}
Here $\alpha_0$ is the fine structure constant, $r_0$ is the classical electron radius, Z is the charge of the nucleus of the absorbing material, and $\beta=v/c$.
The variable $\zeta = \frac{Ze^2}{\hbar\nu}$ and $f(\zeta)$ is unity away  from the absorption edge.

The considerations for the design of a polarimeter that exploits this effect are analogous to those for the scattering polarimeter. 
In this case the competing effects are the desire for a high efficiency for converting the incident X-ray flux into photoelectrons and the desire for those photoelectrons to travel large distances before interacting with elements of the absorbing material. 

Here we will concentrate on polarimeters that use gas mixtures to convert the incident X-rays to photoelectrons.
We do this for the following reasons: (1) there are two promising approaches to electron tracking polarimetry that use this approach and we are quite familiar with both of them; and (2), especially at the X-ray energies of interest here (and where the X-ray fluxes are the greatest), the range of the primary photoelectrons in solids are very tiny (e.g $\simeq$ 1.5$\mu$m in silicon at 10 keV). 
Tracking such events in solids then requires pixels much smaller than the current state of the art, making this type of polarimetry essentially impossible at the energies of interest. 

\begin{figure}
\centerline{\psfig{file=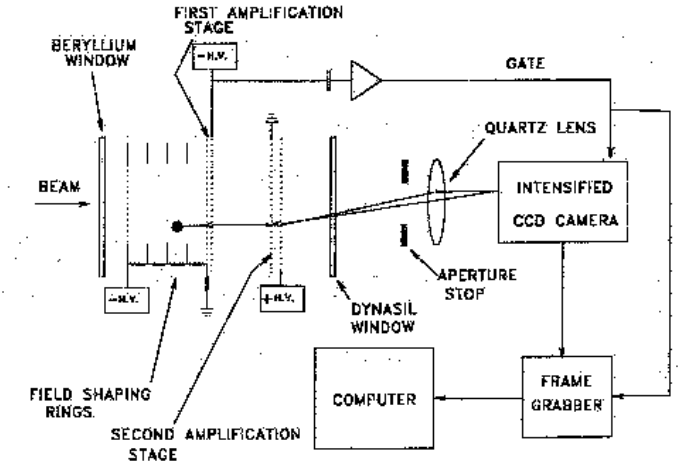,width=9.5 cm,clip=} }
\caption{Experimental setup for the optical imaging chamber.
\label{f:etsetup}}
\end{figure}

To our knowledge, the first electron tracking polarimeter specifically designed to address polarization measurements for X-ray astronomy and using a gas as the photoelectron-emitting material was that designed by Austin \& Ramsey (1992 - see also Austin \& Ramsey 1993; Austin, Minamitami, \& Ramsey 1993) 
These scientists used the light emitted by the electron avalanches which takes place after the release of the initial photoelectron in a parallel plate proportional counter. 
The light was focused and detected by a CCD camera.
A schematic diagram of the experimental setup is shown in Figure~\ref{f:etsetup}.
The use of two multiplication stages (i.e. two parallel-plate proportional chambers) permits triggering of the camera and allows for efficient light yields.
Of course the detection scheme produces a two dimensional projection of the photoelectron's track and this reduces the modulation factor.

\begin{figure}
\centerline{\psfig{file=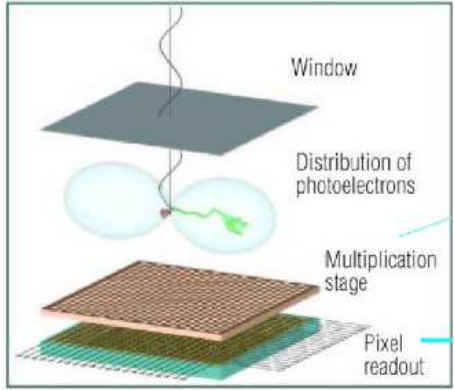,width=6.0 cm,clip=}}
\caption{Cartoon showing the principle of the gas-multiplication electron-tracking polarimeter. Courtesy J. Swank.
\label{f:etgem}}
\end{figure}

Another gas-detector approach, first discussed by Costa et al.\ (2001), uses ``pixillated'' proportional counters to record the avalanche of secondary electrons that result from gas-multiplication in a high field after the the primary photoelectron track (and that of the original Auger electrons) drift into a region where this multiplication may take place. 
The concept is shown in Figure~\ref{f:etgem}.
The resulting charge may then be read out by finely pixellated collection electrodes.

Detecting the direction of the emitted photoelectron (relative to the direction of the incident flux) is itself not simple. 
The reason for this is that electrons, when they interact with matter, give up most of their energy at the {\em end} of their track, not the beginning. 
Of course, in the process of giving up its energy to the local medium in which the initial photo-ionization took place, the electron changes its trajectory, thus losing the information as to the initial polarization.
Therefore, devices that wish to exploit the polarization dependence of the photoelectric effect have the additional challenge that they must track the ejected photoelectron's path, and the most important element of that path is the direction to the first interaction which gives up the least amount of energy. 

\begin{figure}
\centerline{\psfig{file=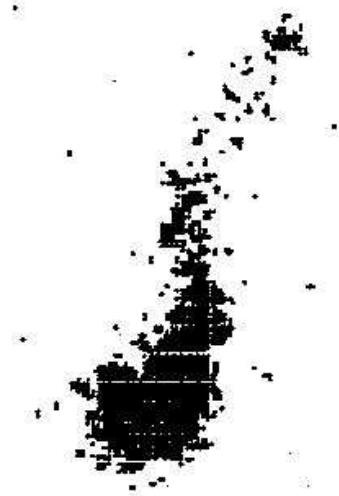,width=6.5 cm,clip=} }
\caption{The two dimensional projection of a track produced when a 54 keV X-ray was absorbed in 2 atm of a mixture of argon(90\%), CH$_4$(5\%), and trimethylamine (5\%). 
The particular track is $\simeq 14$-mm in length. 
\label{f:etrack}}
\end{figure}

It is instructive to examine the image of a track and we show one obtained under relatively favorable conditions with an optical imaging chamber in Figure~\ref{f:etrack}.
The initial photoionization has taken place at the small concentration of light to the north (top) of the figure. 
The bright spot to the north indicates the short track of an Auger electron.
As the photoelectron travels through the gas mixture it either changes direction through elastic scattering and/or both changes direction and loses energy through ionization.
As these take place, the path strays from the direction determined by the incident photon's polarization. 
Of course, the ionization process is energy dependent and most of the electron's energy is lost at the end, not the beginning, of its track.
It should be clear from this picture that, even under favorable conditions  --- by which we mean those where the range of the photoelectron is quite large compared to its interaction length --- the ability to determine a precise angular distribution depends on the capability and sophistication of the track-recognition software, not only the spatial resolution of the detection system.
The burden falls even more heavily on the software at lower energies where the photoelectron track becomes very short and diffusion in the drifting photoelectron cloud conspires to mask the necessary track information.  

Although polarimeters exploiting this effect have been discussed in the literature, {\em no device of this type has ever been flown and those built have undergone limited testing in the laboratory}.
The claims for the potential performance of these devices at the energies of peak performance depend on Monte-Carlo simulations to extend experimental results.
Experimental verification of performance exists at 5.4 and 6.4 keV (Bellazzini et al.\ 2006). 
We eagerly await experimental verification of performance at lower energies, around 3 keV, where peak performance is claimed.

Both approaches for imaging the projection of the electron track are quite interesting, especially for use at the focus of an X-ray telescope (\S \ref{s:wtelescope}). 
Electron tracking polarimeters must also deal with a energy dependent modulation factor.
This is completely in contrast to the crystal polarimeter, and is more severe than for a typical scattering polarimeter.
This energy dependence will not only complicate the calibration of such an instrument, but also the data analysis.
To our knowledge, no published reports of the projected sensitivity of such devices have ever considered the impact of the finite energy resolution and the energy-dependent modulation on the data analysis.
To do so here is beyond the scope of this paper, but we note that the impact of this complication on the sensitivity should not be ignored.

The considerations for the choice of the detector gases are somewhat different for the two approaches to electron tracking discussed here - high light yield versus reasonable electron amplification --- but both must trade a high absorption efficiency for a long electron track in order to work efficiently as a polarimeter.

There are pros and cons in each approach. 
The optical imaging chamber has the advantage of flexibility in its readout scale, which can be configured by the appropriate choice of optics so that its detection pixel is small compared to the electron track length, especially at the low-energy end of the polarimeter response. 
In contrast, the fixed size of the pixels ($\simeq 50-100-\mu$m) themselves determine the low-energy response when gas-multiplication detection is used. 
This is probably more of a limitation than might appear at first sight since the arrangement of detection cells is, in and of itself, asymmetric in position angle, with a built in response at $2\phi$, the signature of polarization. 
This built-in asymmetry not only impacts the modulation factor (it vanishes if the length of the track is smaller than the size of a cell), but also introduces spurious polarization signatures when the track length is comparable to, or even somewhat larger than the characteristic size of a detection cell.
We feel that it is naive to believe that such effects can be accurately accounted for by means of Monte-Carlo simulations alone. 

The optical imaging chamber, however, is more limited in its selection of fill-gas mixtures in that they must produce large amounts of light via the addition of photo-sensitive vapors without any competing (non-light-producing) collisions with other additives. 
This potentially limits control over diffusion which in turn may limit the lowest-energy response of such a device.
More detailed study is required to explore the fill-gas parameter space. 

We encourage all experimenters working with gas-multiplication detectors for use as X-ray polarimeters to publish a calibration using polarized and, equally important, unpolarized sources in the regime for which the range of the photoelectron begins to get even close to the size of the detector pixels or to the diffusion scale so that one may understand the true response.

\subsection{X-Ray polarimeters at the focus of a telescope \label{s:wtelescope}}
We first look at polarimeters at the focus of a telescope, which as we will see, provide the highest sensitivity. 
We shall then turn to polarimeters without telescopes and show that, while producing lower sensitivities, these may still offer the best overall approach for a low-cost pathfinder mission. 

There can be no question that for optimizing signal-to-noise, one should place the X-ray polarimeter at the focus of an X-ray telescope.
Further, the electron tracking devices, if shown to work as advertised, are, probably the instrument of choice.
(We hedge only in that it is unclear that these devices can efficiently work at arbitrary energies, and thus may not be suitable to the study of very soft X-ray sources.)
This stems from the fact that these devices will provide the broadest bandwidth together with a very low background, determined only by the size of the initial ionization convolved with the telescope's angular resolution. 
In contrast, the background for the scattering polarimeter is determined by the area of the surrounding detectors, which, perforce, is much larger.
The background for the crystal polarimeter near the focus of a telescope is also very small as it is determined by the resolution of the X-ray telescope. 
However, the bandwidth is tiny, unless multiple crystals are utilized. 
A multiple-crystal design is complex, and beyond the scope of this discussion. 
(Possible a hybrid with thin crystals operating in series with an electron-tracking device might be interesting.)

In order to perform a comparison with the same telescope for all three types of instruments, and to make use of existing software, we consider a graphite-crystal polarimeter, a lithium-scattering polarimeter, and a photoelectron tracking polarimeter, each at the focus of the SODART telescope.
This was a 60-cm-diameter, 8-m-focal-length, foil telescope of $\simeq$ 1000 $cm^{2}$ at 3 keV built for the (original) Spectrum-X mission.

\begin{figure}
\centerline{\psfig{file=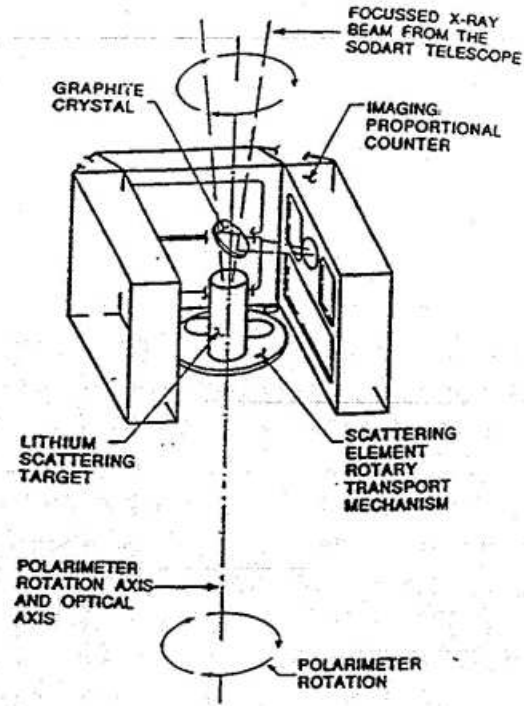,width=8.5 cm,clip=} }
\caption{Cartoon showing the Stellar X-Ray Polarimeter built for Spectrum-X.
\label{f:sxrp}}
\end{figure}

The configurations we consider are as follows: a graphite-crystal polarimeter followed by a lithium-scattering block surrounded by a four-proportional-counter array, as were employed for the Stellar X-Ray Polarimeter (SXRP - see Fig.~\ref{f:sxrp} and also Kaaret et al.\ 1994 and references therein) built for the (original) Spectrum-X mission, and an electron tracking polarimeter filled with a mixture of 80\% Ne, 20\% DME at 1 atm, with a 100-micron-pitch readout, as simulated by Pacciani et al.\ 2003.

\begin{figure}
\centerline{\psfig{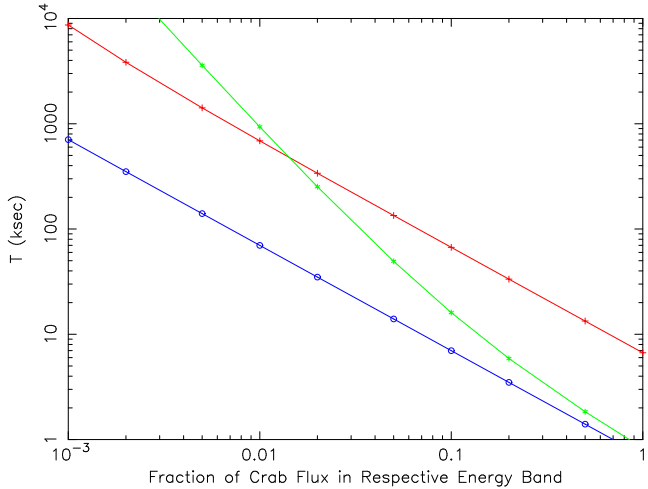}}
\caption{Times to reach a MDP of 3\% versus source strength for the Crab Nebula spectrum. 
The numbers result from integrating over useful energy response of each instrument,
The blue line is for an electron tracking polarimeter, the green is for a lithium scattering polarimeter, and the red for a graphite crystal polarimeter. 
Note that the latter two operate simultaneously. 
All placed are placed at the focus of the SODART telescope.
\label{f:crabtvsf}}
\end{figure}

\begin{figure}
\centerline{\psfig{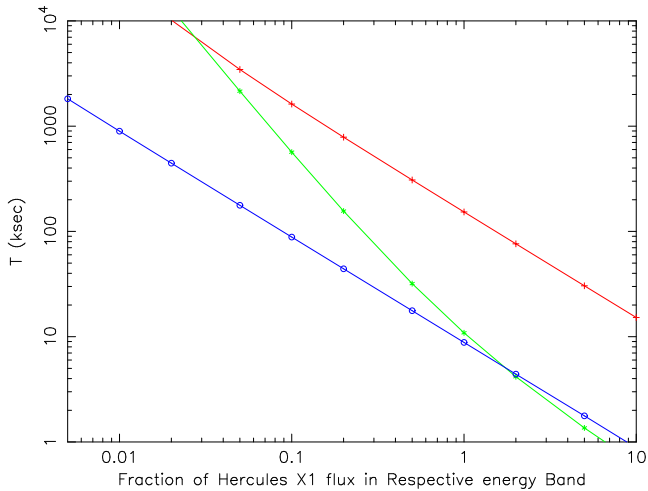}}
\caption{Same as for Figure~\ref{f:crabtvsf} but for the Her X-1 spectrum.
\label{f:hertvsf}}
\end{figure}

Figures~\ref{f:crabtvsf} \& \ref{f:hertvsf} compare the times to reach 3\% MDP (at the 99\%-confidence level) for the graphite-crystal, lithium-scattering block, and the electron-tracking polarimeter, in various energy bands for two different incident energy spectra.
The calculations for the Spectrum-X polarimeters are based on Monte-Carlo simulations fully verified by calibration measurements at LLNL (Silver et al.\ 1994), while those for the electron-tracking polarimeter are based on data taken from published simulations (Pacciani et al.\ 2003). 
The graphite-crystal and electron-tracking polarimeters are not background limited, at least down to source strengths corresponding to a milliCrab, while the lithium-scattering block polarimeter is background limited over the entire range of source strengths shown.
Since the electron-tracking polarimeter is more sensitive to fainter sources, it seems clear that {\em all things being equal} one would choose to place the electron-tracking polarimeter at the focus of an X-ray telescope especially if one had to choose single device.

In performing these calculations and the comparisons, we have ignored the specter of systematic effects that might lead to false signatures of polarization, and hence reduced sensitivity.
Accounting for such effects is of great importance, especially at low levels of polarization which are exacerbated by below-unity modulation factors.
{\em With all deference to high-fidelity Monte-Carlo simulations, careful ground-based calibrations over the entire operating range of a polarimeter, performed with both polarized and unpolarized beams are essential for establishing performance.} 
The more complex the polarimeter, the more important such calibrations become.
Frankly, the literature has several examples of highly exciting new approaches to polarimetry, which, on deeper experimental examination have turned out to be incorrect and dominated by systematic effects (see, e.g., Shaw et al.\ 1991).

The systematic effects may not be limited to the polarimeter itself.
Items that also need to be considered are, e.g., the coupling of the telescope to the instrument --- especially if the telescope is not round; off-axis effects (see Elsner et al.\ 1990 for one of these effects for scattering polarimeters); and solar X-rays that have become polarized through scattering through the appropriate angles from the atmosphere.

\subsection{X-Ray polarimeters without a telescope\label{s:ntelescope}}

The comparison amongst the three approaches to X-ray polarimetry we are considering here is quite different if we examine polarimeters without telescopes. 
Now the devices that track the photoelectron, so useful at the focus of the telescope, are no longer really practical because of the large detector area and small pixel size (to establish the photoelectron track) that are both required. Thus, we examine the question how best to fill a modest aperture with a polarimeter that does not involve an X-ray telescope and in this context compare large-area scattering and crystal polarimeters.

\begin{figure}
\centerline{\psfig{file=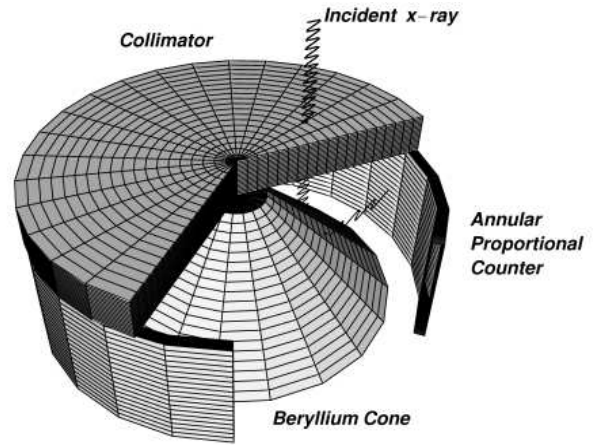,width=9.0 cm,clip=} }
\caption{Conceptual design for a beryllium-scattering polarimeter.
\label{f:bepol}}
\end{figure}

For the purpose of this comparison we consider a beryllium scattering polarimeter (XPE) which is a realization of a design we first introduced in cartoon form in M\.{e}sz\'{a}ros et al.\ (1988). 
The design is illustrated in Figure~\ref{f:bepol} and consists of a 0.6-m-diameter beryllium scattering cone surrounded by an annular proportional counter to record the angle and energy of scattered photons. 
A simple collimator limits the field of view to a few degrees.
Note that the diameter of the opening is identical to that of the SODART telescope we used with the polarimeters in \S \ref{s:wtelescope}, thus the filling-factors are identical. 

\begin{figure}
\centerline{\psfig{file=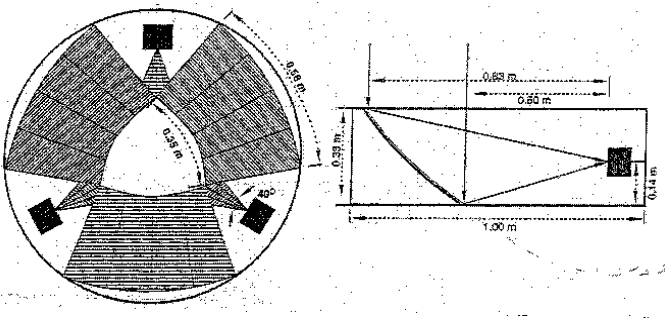,width=9.0 cm,clip=} }
\caption{Conceptual design for a crystal polarimeter (PLEXAS).
\label{f:plexas}}
\end{figure}

For a typical large-area crystal polarimeter we consider an array of multilayer-coated reflectors tuned for high throughput at large graze angles (25-40 degrees) at 0.25 keV This is the PLEXAS design of Marshall et al.\ 2003.
The reflectors are arrayed in three sectors, each sector reflecting onto a different detector.
The concept is illustrated in Figure~\ref{f:plexas}.
The footprint of both polarimeters is also similar.

\begin{figure}
\centerline{\psfig{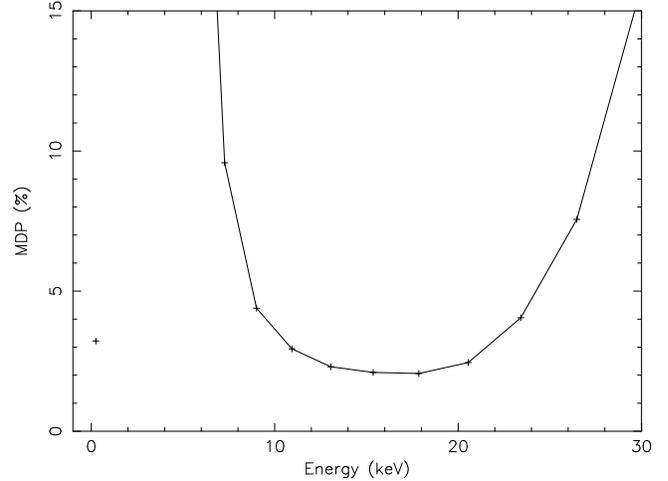}}
\caption{The minimal detectable polarization for a $10^5$ sec integration versus energy for the two polarimeters without a telescope.
The single point at 0.25 keV is for the synthetic multilayer (PLEXAS) design.
The continuous line is for the beryllium-scattering polarimeter. 
\label{f:noteleherx1}}
\end{figure}

\begin{figure}
\centerline{\psfig{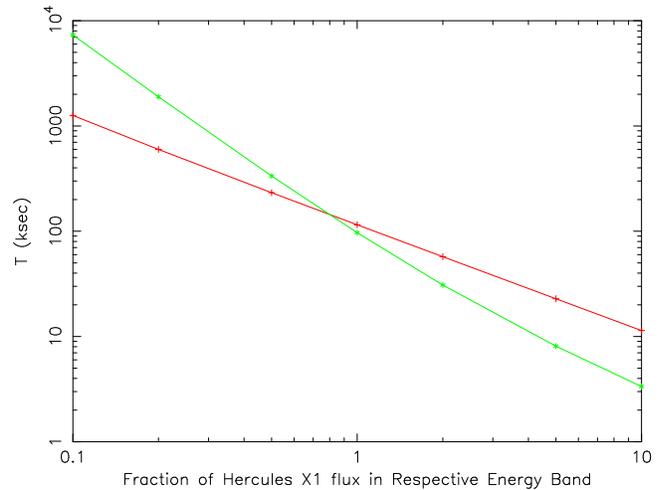}}
\caption{The integration time T to reach a MDP of 3\% (at 99\%-confidence) versus source strength for the Her X-1 spectrum integrated over the full energy response of each polarimeter.
Note that the bandwidths are quite different.
The green line is for the collimated beryllium-scattering polarimeter. 
The red line is for the synthetic multilayer operating nominally at 0.25 keV.
\label{f:notelecompare}}
\end{figure}

Figure~\ref{f:noteleherx1} shows polarization sensitivity for Her X-1 as a function of energy for the two polarimeters.
The scattering polarimeter achieves peak performance at higher energies than the polarimeters at the focus of the long-focal-length X-ray telescope.
Figure~\ref{f:notelecompare} shows the time to reach 3\% MDP at 99\%-confidence with such polarimeters versus source strength for the Her X-1 spectrum.
Now, in contrast to Figure~\ref{f:noteleherx1}, we integrate over the full bandwidth.

Although, by virtue of its concentrating reflectors, the integrated performance of the synthetic crystal polarimeter is superior for faint sources, it lacks broad band response, and one needs to answer the question as to whether or not the measurement of polarization at a single energy is capable of providing useful additional constraints of our understanding of astrophysical systems.
We strongly believe that the answer to this question is no --- that even a detection at a single energy is not terribly useful. 
In such cases we feel that the clever theorist will soon provide a myriad of ex-post-facto models to explain any unexpected result, and the ability to distinguish between models will be missing.

It should be clear then that without a telescope, a scattering polarimeter is the instrument of choice as it provides useful sensitivity over a wide energy band, especially when compared to the use of a single crystal material or an ungraded multilayer reflector.

\section{Discussion and conclusions\label{s:conclusion}}

There are no free rides in X-ray polarimetry:
An instrument with some polarization sensitivity, but designed primarily for other purposes, is not an adequate substitute for one optimized for polarimetry.
For example, attempts (Coburn \& Boggs 2003; then Rutledge \& Fox 2004) to measure the polarization of GRB 021206 using the Reuven Ramaty High-Energy Solar Spectroscopic Imager (RHESSI) led to results that are controversial at best.
In this case, the low priority for possible polarization measurements practically precluded the complete calibration needed to characterize an instrument's polarimetric sensitivity and to understand systematic effects that might produce a spurious polarization signal.
For instruments operating at high energies, such a calibration could require exposing the entire spacecraft to an X-ray beam and would thus be difficult.
While Monte-Carlo simulations play an important role in assessing an instrument's capabilities for polarimetry, verifying the quantitative predictions of such simulations still requires careful comparison with calibration or other experimental data.

It is reasonable to ask, ``Why has there been no X-ray polarimetry of cosmic sources since the early experiments in the 1970s?''
Understandably, the development of X-ray astronomy has focused on X-ray optics for imaging and spectrometric imaging, leading to facility-class missions serving a broad astronomical community.
Focused imaging greatly increases the signal-to-noise ratio and mitigates source confusion, dramatically improving sensitivity and thus enabling meaningful observations of a large number of sources, their spectra, and (for resolved sources) their spatial structure.
The {\sl Einstein} Observatory evolved into the first of the facility-class X-ray missions and (unfortunately) became a paradigm for polarimetry in such missions:
The original design for the {\sl Einstein} Observatory (nee HEAO-B) included a polarimeter; however, program restructuring and descoping deleted the instrument.
Although the \cha\ (nee AXAF) call for instruments did not preclude a polarimeter, imaging and spectroscopic instruments prevailed~--- in large part, because many more targets are accessible to such observations than to polarimetry.
In the exceptional case of Spectrum-X, which included a polarimeter insertable into the focal position, competition with the other insertable focal-plane instruments resulted in an observing plan that would have limited polarimetry to only 11 days per year!
Even worse, that observatory never flew!

The absence of any X-ray polarimetric observations since the original experiments has itself impaired the development of X-ray polarimetry.
Without experimental results or even the prospect thereof, progress in the theoretical framework that such experiments both require and inspire has~--- with notable exceptions (\S \ref{s:science})~--- been slow.
We hasten to add that this does not mean an absence of theoretical interest.
Indeed, the 2004 conference on X-ray polarimetry at SLAC \footnote{http://www-conf.slac.stanford.edu/xray\_polar/talks.html}
attracted over 100 scientists, the majority of whom are theorists.

Such considerations have convinced us that a small, dedicated mission affords the best opportunity for advancing X-ray polarimetry.
This permits formulation of an observing program suited to the capability of the polarimeter and avoids the limitations that a shared (e.g., facility-class) mission imposes on the least sensitive instrument aboard.
Even so, it is extremely difficult~--- once again for many of the reasons discussed above~--- for a polarimetry mission to compete with other missions (most outside X-ray astronomy) seeking similar resources (e.g., in the NASA's Small Explorer Program).

Consequently, we believe that an X-ray-polarimetry pathfinder needs to be an inexpensive, simple instrument, with minimal technical requirements upon the spacecraft~--- e.g., pointing accuracy and stability (Elsner et al.\ 1990)~--- and upon the launcher.
Regrettably, such budgetary constraints probably preclude use of a focusing X-ray telescope on the pathfinder.
Suitable X-ray optics are costly to design and fabricate, align and assemble, integrate, and test and calibrate.
Further, even lightweight optics would burden the weight budget for a small spacecraft, especially for a telescope optimized for the higher X-ray energies at which the scientifically more interesting polarization effects (\S \ref{s:science}) are likely to occur.

In view of these constraints, we propose an initial exploratory polarimetry mission, to survey bright X-ray sources, using a large-area scattering polarimeter, possibly supplemented with crystals.
This type of instrument is simple (no deployables or other moving parts), low-cost, and proven.
We estimate that the instrument costs would be around 5~M\$ and that the total mission cost would be about 30~M\$.
This is roughly a quarter of the cost of the typical NASA Small Explorer program, where fixed prices for complex 3-axis-stabilized catalog satellites, large launch costs, etc.\ mask the true cost of a simple mission.

Such a pathfinder could survey a wide range of objects at sufficient sensitivity to detect expected levels of polarization.
To illustrate this, Table~\ref{t:mdp} lists a sample survey program, with the integration time and MDP for the XPE polarimeter (\S \ref{s:ntelescope}).
Each integration time is that necessary to yield 3\% MDP (integrated over energy and phase, if pulsating) or 0.5 days, whichever is longer.
After the 6 months needed to complete the survey (including time for slewing and target acquisition), the remainder of the mission would conduct follow-on measurements of many of the sources exhibiting a polarization signature.
In addition to performing the first X-ray-polarimetry survey, the low-cost pathfinder would serve as the foundation for a larger, more-complex mission that could include large-area focusing optics with fully developed and calibrated electron-tracking polarimeters at their foci.

\begin{table}
\caption{Sample Polarimetry Survey\label{t:mdp}}
\begin{tabular}{|l|l|c|c|}
\hline 
Name         & Type                   & Time   & MDP\\
             &                        & (d)    & (\%)$^a$\\ \hline
Crab Pulsar  & Radio Pulsar           & 29.6   & 3.0\\
Crab Nebula  & SNR                    &        & 0.1\\
SGR 1900+14  & SGR: in active state   & 1      & 3.0 \\
4U1636-53    & burster                & 9      & 3.0\\
GS1826-238   & Clocked burster$^b$    & 4.3    & 3.0\\
J1808.4-3658 & MSP                    & 9.1    & 3.0\\
J1751-305    & MSP                    & 10.3   & 3.0\\
Her X-1      & Accreting pulsar       & 0.5    & 1.9\\
Cen X-3      & Accreting Pulsar       & 0.5    & 1.4\\
4U0900-40    & Accreting Pulsar       & 0.5    & 2.4\\
GX 1+4       & Accreting Pulsar       & 0.5    & 2.1\\
SMC X-1      & Accreting Pulsar       & 3.2    & 3.0\\
4U1538-58    & Accreting Pulsar       & 10.4   & 3.0\\
4U0115+63    & Accreting Pulsar$^c$   & 0.5    & 2.4\\
OAO1657-41   & Accreting Pulsar       & 4      & 3.0\\
4U1626-67    & Accreting Pulsar       & 1.3    & 3.0\\
Cyg X-3      & Binary                 & 1.0    & 3.0\\
4U1822-37    & Accretion-Disk Corona  & 8.3    & 3.0\\
Sco X-1      & QPO                    & 0.5    & 0.6\\
Cyg X-2      & QPO                    & 0.5    & 2.8\\
GX 5-1       & QPO                    & 0.5    & 2.0\\
Cir X-1      & QPO                    & 0.5    & 2.0\\
Cyg X-1      & Black-hole binary      & 0.5    & 0.9\\
J1744-28     & Bursting pulsar        & 0.5    & 0.6\\
GRS 1915+105 & Microquasar$^d$        & 0.5    & 0.4\\
J1655-40     & Microquasar$^c$        & 0.5    & 1.8\\
TBD          & Weak Transient         & 10.7   & 3.0\\
Cen A        & AGN                    & 16.2   & 3.0\\
NGC 4151     & AGN                    & 24.8   & 3.0\\ \hline
\end{tabular}
$^a$~99\% confidence \ 
$^b$~persistent flux \ 
$^c$~high state \ 
$^d$~active state
 \end{table}

\vskip 0.4cm

\begin{acknowledgements}

MCW gratefully acknowledges support by NASA, by the Heraeus Foundation, and by the Max-Planck-Institut f\"ur extraterrestriche Physik in Garching.
\end{acknowledgements}

\end{document}